# Visco-elastic drag forces and crossover from no-slip to slip boundary conditions for flow near air-water interfaces

A. Maali[*], R. Boisgard, H. Chraibi, Z. Zhang, H. Kellay and A. Würger

Université de Bordeaux & CNRS, LOMA, UMR 5798, F-33400 Talence, France

The "free" water surface is generally prone to contamination with surface impurities be they surfactants, particles or other surface active agents. The presence of such impurities can modify flow boundary near such interfaces in a drastic manner. Here we show that vibrating a small sphere mounted on an AFM cantilever near a gas bubble immersed in water, is an excellent probe of surface contamination. Both *viscous* and *elastic* forces are exerted by an air-water interface on the vibrating sphere even when very low doses of contaminants are present. The viscous drag forces show a cross-over from no-slip to slip boundary conditions while the elastic forces show a nontrivial variation as the vibration frequency changes. We provide a simple model to rationalize these results and propose a simple way of evaluating the concentration of such surface impurities.

* Corresponding author : abdelhamid.maali@u-bordeaux.fr

Micro-scale and nano-scale flows near fluid interfaces are very sensitive to the presence of surface active material. On the one hand, surfactant molecules give rise to a surface shear viscosity which has been observed through the damping of surface waves [1-2], the enhancement of the drag coefficient of floating beads [3] or disks [4], and the self-propulsion velocity of colloidal micro-swimmers [5]. On the other hand, a dilute surfactant monolayer reduces the interface tension by the ideal-gas pressure $\Pi = ck_BT$, where $k_BT$ is the thermal energy. A non-uniform surface flow locally modifies the concentration $c$; and its gradient $\nabla c$ gives rise to a tangential stress $k_BT\nabla c$ which, in turn, induces new properties such as dilatational viscosity and surface elasticity [6-7]. In many instances, the Marangoni effect resulting from the surface tension gradient completely changes the flow pattern: It may reduce the liquid slip on superhydrophobic surfaces [8-9], and change the flow profile in thin films as observed through the measurement of the hydrodynamic interaction between two bubbles using an atomic force microscope (AFM) [10-12].

Confined flows are to a large extent determined by the hydrodynamic boundary conditions imposed by material properties, such as the hydrophobicity of solid surfaces [13-18] and the molecular structure [19]. An efficient tool for the study of flows near boundaries is a colloidal sphere moving towards a flat surface. Close to a no-slip surface, the drag force exerted on the sphere reads [16]

$$F_0 = -6\pi\eta R_{eff}^2 V/d, \quad (1)$$

where $\eta$ is the viscosity of the fluid, $R_{eff}$ the hydrodynamic radius of the sphere, $V$ its velocity, and $d$ the distance between the bottom of the sphere and the surface. Close to a free surface with full slip, however, the drag force is four times smaller and is given by $F_0/4$ [17]; a similar effect occurs for moving air bubbles [20]. Slip at solid-gas interfaces depends both on kinetic parameters [21] and adsorption [22].

At first sight, an air-water interface is expected to behave as a free surface. However, a recent experimental study [10-11] showed that the full-slip condition is not realized in general; the measured drag force rather corresponds to an intermediate situation. As a possible explanation for this increase of viscous forces for the 'bare' water surface, the authors invoked the presence of impurities.

In this Letter, we show that dynamic AFM measurements coupled to a model based on lubrication theory with appropriate boundary conditions (accounting for residual Marangoni stresses due to the presence of minute amounts of impurities), can account for the role of such impurities at a 'neat' water surface. The dynamic AFM mode with different operating frequencies allows for a direct measurement of both elastic and viscous drag forces and their modeling gives access to the concentration of contaminants at the surface. Even minute amounts of such contaminants modify boundary conditions in a dramatic way giving rise to a crossover from no-slip to full slip conditions when increasing the solicitation frequency.

A sketch of the experimental configuration is shown in Fig. 1. The air-water interface was prepared by deposing a small gas bubble on a polystyrene covered glass surface. A glass sphere glued at the end of an AFM cantilever which is used to induce oscillations and to measure the resulting hydrodynamic drag force on the sphere.

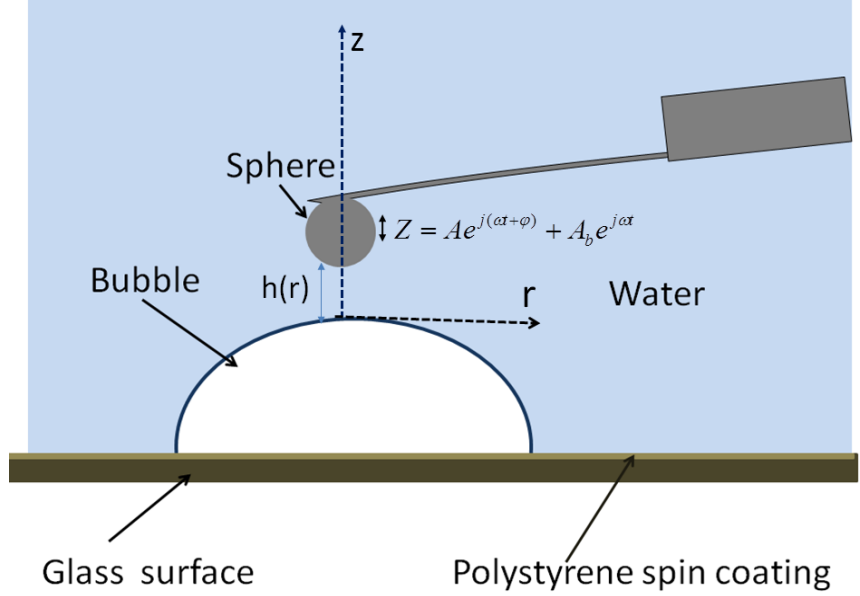

**Fig.1**: sketch of the experimental apparatus. The liquid/gas interface was prepared by deposing a spherical bubble on PS surface. A vibrating glass sphere glued at the end of the AFM cantilever is used to induce the hydrodynamic drag force.

In order to keep the analysis of the measurements simple, we work in the range of parameters where the capillary deformation $u$ of the bubble is small as compared to the amplitude $V/\omega$ of the cantilever vibration. The oscillation velocity $V$ results in a drag force $F_0$ and in a deformation field $u \approx F_0/\sigma$, where $\sigma$ is the interface tension. One readily finds that capillary effects are irrelevant as long as the tip-bubble distance obeys $d > 6\pi\eta R^2 \omega/\sigma$. With $R \sim 50$ µm, $\omega \sim 10^3$ rad s$^{-1}$ and $\sigma \sim 72.8$ mN/m, this condition gives $d > 0.5$ µm.

For a cantilever excited at frequency $\omega$, the displacement motion of the tip $z(t)$ is described by the oscillator model equation

$$m^*\ddot{z} + \Gamma_{bulk}\dot{z} + k_c z = F_{drive} + F_h, \qquad (2)$$

with the effective mass of the cantilever $m^*$, the force constant $k_c$, and the bulk damping coefficient $\Gamma_{bulk}$. The forces acting on the sphere are the driving $F_{drive}$ [23] and the hydrodynamic drag $F_h$ resulting from the interaction with the interface.

In the standard dynamic AFM where the driving force is induced by the cantilever base displacement [23-24], the measured signal is the oscillation of the cantilever deflection $Ae^{j(\omega t+\varphi)}$, whereas the position of the tip $z(t)$ includes the cantilever base displacement with amplitude $A_b$ [23-24]: $z(t) = Ae^{j(\omega t+\varphi)} + A_b e^{j\omega t}$.

The steady-state solution $z(t)$ of equation (2) gives rise to a linear relation between the tip velocity $\dot{z}$ and the hydrodynamic force,

$$F_h = -\frac{k_c(1-(\frac{\omega}{\omega_r})^2 + j\frac{\omega}{\omega_r Q})}{-j\omega}(\frac{Ae^{j\varphi} - A_\infty e^{j\varphi_\infty}}{Ae^{j\varphi} + A_b})\dot{z} \qquad (3)$$

The viscoelastic response could be equally well described in terms of a complex hydrodynamic drag coefficient $\Gamma_h(\omega) = \Gamma_{vis} - j\Gamma_{el}$ which is defined through $\Gamma_h = -F_h/\dot{z}$. $A_\infty$ and $\varphi_\infty$ are respectively the free amplitude and phase measured far from the surface. Moreover we have defined the angular resonance frequency of the cantilever far from the surface $\omega_r = \sqrt{k_c/m^*}$ and the bulk quality factor of the cantilever $Q = m^*\omega_r/\Gamma_{bulk}$. At a given frequency, the hydrodynamic drag force, and thus the real and imaginary parts of the hydrodynamic drag coefficient are determined from the measured variation of the amplitude and phase $A$ and $\varphi$, using the parameters $\omega_r, Q, A_\infty, \varphi_\infty$ of the resonance spectrum of the cantilever far from the surface [23].

The experiment was performed using an AFM (Bruker, Bioscope) equipped with a liquid cell (DTFML-DD-HE) that allows tapping mode in a liquid environment. We have used a spherical borosilicate particle (MO-Sci Corporation) of radius $R = 53.1\,\mu m$. The sphere (cleaned using ethanol and pure water) was glued to the end of a silicon nitride rectangular cantilever ORC8 (Olympus) using epoxy (Araldite, Bostik, Coubert). The assembly of sphere and cantilever was then rinsed several times with ultrapure water. The water is obtained from a MilliQ-Millipore ultrapure water system. The liquid cell was cleaned using ethanol and rinsed several times with pure water. The samples studied in this paper were fixed on a multi-axis piezo-system (NanoT series, Mad City Labs ) that allows a large displacement (up to 50µm) with a high accuracy under closed loop control. Using the drainage method described by Craig et al [25], the stiffness of the cantilever with an attached sphere was determined from the drainage data at large enough distances (200-30000nm), and was found to be $k_c = 0.249 N/m$. The bubble radius measured with an optical microscope is $R_b = 378.4\,\mu m$, resulting in an effective hydrodynamic radius on the bubble $R_{eff} = 1/(R_b^{-1} + R^{-1}) = 46.6\,\mu m$.

The cantilever was vibrated at fixed frequency; the amplitude and phase of the cantilever were recorded while the sphere approached the interface at velocities smaller than 0.4 µm/s. The DC deflection was also recorded and used to determine the tip position [23]. The quality factor of the cantilever in bulk water is $Q = 3.9$, and the resonance occurs at $\omega_r / 2\pi = 1340\ Hz$.

In order to verify the sensitivity of our apparatus, we show in Fig. 2a the viscous and the elastic parts of the hydrodynamic drag coefficient on a sphere vibrating close to a mica surface. As expected, the hydrodynamic interaction with the mica surface is purely viscous. Fig. 2b shows the hydrodynamic drag coefficient versus the distance for different vibration frequencies. Note that the hydrodynamic drag coefficients extracted for different vibration frequencies coincide with each other, and with the theoretical hydrodynamic drag coefficient $6\pi\eta R_{eff}^2 / d$ for no-slip boundary conditions in agreement with previous results [15][18].

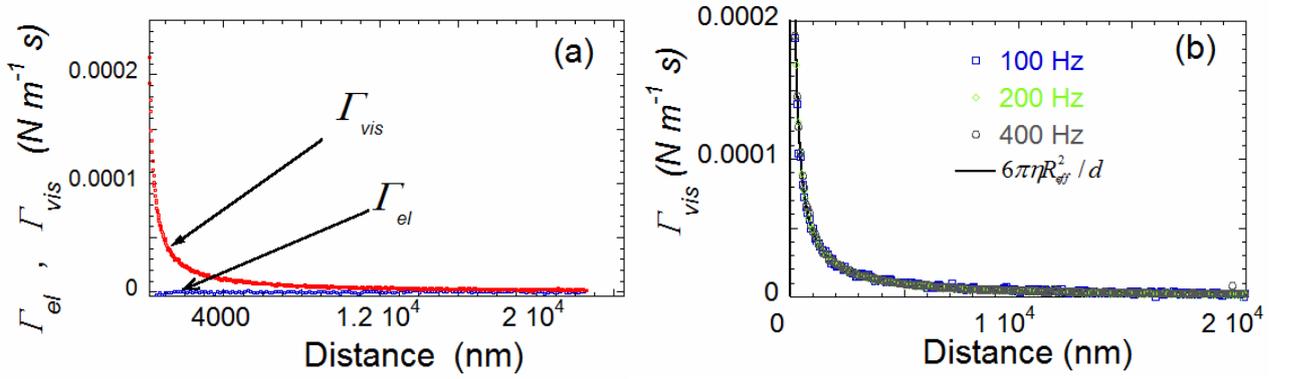

**Fig.2**: (a) shows the viscous $\Gamma_{vis}$ and the elastic part $\Gamma_{el}$ of the hydrodynamic drag coefficient for the sphere vibrating at frequency (200 Hz) in water close to a mica surface. (b) The hydrodynamic drag coefficient versus the distance for different vibration frequencies. The solid dark line is the theoretical drag coefficient $\Gamma_0 = 6\pi\eta R^2 / d$ calculated with no-slip boundary condition on both surfaces (glass sphere and mica substrate).

Fig. 3 shows the drag coefficient measured on the bubble. Unlike the measurements on the mica surface, the results show that the interaction is not purely viscous (Fig.3a). This viscoelastic response contains two measurable components, viscous and elastic. Further, the viscous component for different frequencies of vibration do not coincide with each other as for mica (Fig.3b). In this figure, the drag coefficients corresponding to full-slip and no-slip boundary conditions on the bubble surface are shown. While for low frequencies, the viscous drag is close to the no-slip case, with increasing frequency, the drag force decreases and finally approaches the full slip boundary condition on the bubble. In our experiment, the frequency could not be increased further since vibrations in bubble shape are excited at higher frequencies. (For a $400\ \mu m$ bubble, the first resonance occurs around 600 Hz [26].)

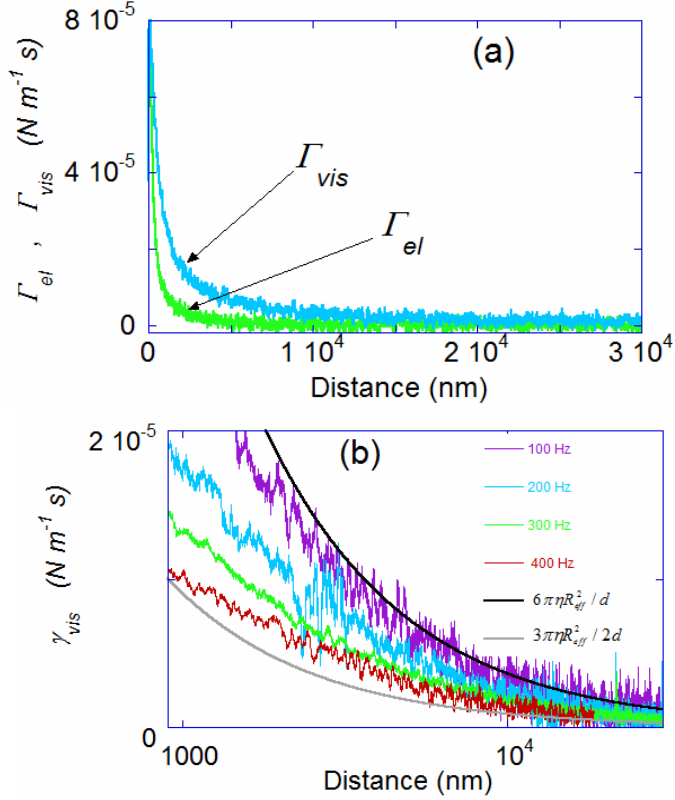

**Fig.3**: (a) The two components of the drag coefficient measured on the bubble surface (sphere vibration frequency 100 Hz). (b) Viscous component for different frequencies of vibrations. The calculated drag coefficient corresponding to full slip and no-slip boundary conditions on the bubble surface are represented by the grey and dark line respectively.

In order to rationalize our data on the bubble surface we assume that the air-water interface contains surface impurities as has been suggested previously [8-11]. In general, surface impurities are characterized by a surface concentration $c$ which obeys an advection-diffusion equation [10-11],

$$\frac{dc}{dt} + \nabla \cdot (v_s c) = D\nabla^2 c, \quad (4)$$

where $D$ is the surface diffusion coefficient. For symmetry reasons, the 2D gradient operator $\nabla$ has a radial component only. The advection term arises from the radial velocity $v_r$ at the bubble surface, $v_s = v_r(z=0)$.

The velocity field in the water film between the bubble and the bead is treated in standard lubrication theory, with the thickness $h(r) = d + r^2/2R_{eff}$. The boundary conditions for the vertical and radial velocity components read $v_z(z=0) = 0$ and $\eta \partial_z v_r(z=0) = \partial_r \Pi$. Here $\Pi$ is the surface pressure resulting from the presence of surface active impurities. One finds that both the surface velocity

$$v_s = v_s^0 - \frac{h}{4\eta}\frac{\partial \Pi}{\partial r} = -\frac{3rV}{4h} - \frac{h}{4\eta}\frac{\partial \Pi}{\partial r} \quad (5)$$

and the gradient of the hydrodynamic pressure

$$\frac{\partial P}{\partial r} = \frac{3}{2}\left(\frac{\eta r V}{h^3} - \frac{1}{h}\frac{\partial \Pi}{\partial r}\right) \qquad (6)$$

depend on the tangential stress $\partial \Pi / \partial r$ caused by the impurities [10-12]. Eq. (5) describes the back-reaction of the impurity concentration profile on the hydrodynamic flow. For dense monolayers, $\Pi$ is an intricate function of concentration, whereas in the present case of dilute residual impurities, the surface pressure follows the ideal-gas law $\Pi = c k_B T$. With the pressure $P(r)$ calculated from (6), one obtains the hydrodynamic force on the sphere as a surface integral:
$$F_h = 2\pi \int_0^\infty p(r)\, r\, dr, \qquad (7)$$

In our experiments the diffusion term in (4) is irrelevant. Indeed, with $R_{eff} \approx 50\,\mu m$ µm, $d = 1\ldots 10$ µm, and $D \sim 10^{-10}$ m²/s, one finds that the time of diffusion $\tau = L^2/D$ over the lubrication length $L = \sqrt{2 R_{eff} d}$, by far exceeds the period of the cantilever vibration. For weakly soluble impurities, the characteristic relaxation time involves diffusion in the aqueous phase over the distance $L$, resulting in $\tau_b = L^2/D_b$ with the bulk diffusion coefficient $D_b$. For the range of oscillation frequencies used in the experiment, both $\omega \tau$ and $\omega \tau_b$ are significantly smaller than unity. Thus impurity diffusion along the interface, or in the thin water film, is slow as compared to the advection, and may be discarded in the equation of motion (4).

By writing $\nabla \cdot (v_s c) = v_s \cdot \nabla c + c \nabla \cdot v_s$ and neglecting small terms proportional to the concentration gradient $\nabla c$ or to the concentration modulation with respect to the equilibrium value $c_0$, Eq. (4) simplifies to

$$\frac{dc}{dt} - \frac{c_0 k_B T}{4\eta}\frac{\partial}{r \partial r}(r h \frac{\partial c}{\partial r}) = c_0 \frac{3 V d}{2 h^2}, \qquad (8)$$

where the second term on the left-hand side arises from $\nabla \Pi$, and the right-hand side from the divergence of the unperturbed surface velocity $v_s^0$.

The concentration modulation due to the second term is in phase with the oscillatory velocity $V$, whereas the first one is out of phase, as is clear from the ansatz $dc/dt = j\omega c$. The in-phase term with the velocity modifies the prefactor of the viscous drag exerted on the vibrating sphere; the out-of-phase results in an elastic response which is absent at a free surface. The relative weight of surfactant-induced viscous and elastic forces is given by the ratio of the driving frequency $\omega$ and the parameter

$$\omega_0 = \frac{c_0 k_B T}{8 \eta R_{eff}}, \qquad (9)$$

which is obtained by replacing *h* with *d*, and the gradients in the second term of (8) with the inverse lubrication length $\sqrt{2R_{eff}d}$.

Using a finite-element method, Eq. (8) is solved numerically to get the value of the surface pressure $\Pi = ck_BT$ and the hydrodynamic drag force. For the two limiting cases $\omega \ll \omega_0$ and $\omega \gg \omega_0$, the expression of the drag force can be calculated analytically.

In the quasi-static limit, at small frequency or high impurity concentration, we treat *dc/dt* in (8) as a perturbation. Solving the differential equation to linear order in the small parameter $\omega/\omega_0$ by iteration, and integrating (6) and (7) with the resulting *c(t)*, we find

$$F_{vis} = F_0 \text{ and } F_{el} = F_0 \frac{3\omega}{8\omega_0} \quad , \quad (\omega \ll \omega_0) \quad . \quad (10)$$

with the reference force $F_0$ given by Eq.(1). The viscous drag corresponds to that on a solid surface with no-slip boundary conditions, which is four times larger than on a free surface [16]; in physical terms this arises from the surfactant-induced surface stress $\nabla\Pi = k_BT\nabla c$ and its back-reaction on the surface flow.

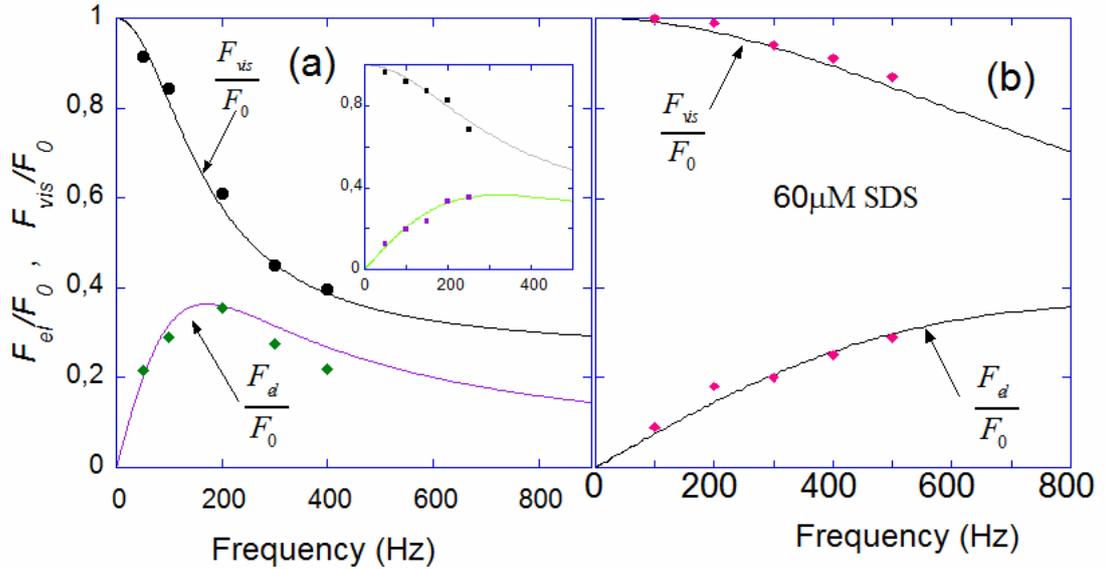

**Fig 4:** (a) $F_{vis}/F_0$ and $F_{el}/F_0$ versus the vibration frequency, for the data of a first experiment. The data of a second run, obtain one month later under similar conditions, are shown in the inset. The theory (solid line) agrees very with the data, with fitted equilibrium surfactant pressure $\Pi_0 = 0.25$ mN/m and $\Pi_0 = 0.35$ mN/m respectively. (b) Data of a control experiment on a 60 μM SDS solution fitted with a surface pressure $\Pi_0 = 1.0$ mN/m.

On the contrary, for sufficiently low impurity concentration and high frequency, $\omega \gg \omega_0$, the second term on the left-hand side in (8) can be treated as a perturbation. Evaluating to second order and integrating Eqs. (6) and (7), we obtain

$$F_{vis} = F_0 (\frac{1}{4} + 8\frac{\omega_0^2}{\omega^2}) \text{ and } F_{el} = F_0 \frac{2\omega_0}{\omega} \quad (\omega \gg \omega_0). \quad (11)$$

In this limit the surface stress $k_B T \nabla c$ is of less importance, and for $\omega/\omega_0 \to \infty$ we recover the drag force for perfect slip, $F_{vis} = F_0/4$. The elastic force varies linearly with $\omega_0/\omega$.

We have performed two independent experiments under similar conditions, at a temporal distance of one month. In Fig. 4a we present the measured viscous and elastic drag force divided by the reference force (1), and we compare with numerical calculations (solid line). The only adjustable parameter is the impurity concentration $c_0$ that defines the surface pressure $\Pi_0 = c_0 k_B T$. The fitted values are $\Pi_0 = (0.25 \pm 0.05)$ mN/m for the first run and $(0.35 \pm 0.05)$ mN/m for the second one, corresponding to $c_0 = (63 \pm 13) \times 10^{15} m^{-2}$ and $(87 \pm 13) \times 10^{15} m^{-2}$, or to an area per molecule of $16\ nm^2$ and $12\ nm^2$, thus justifying the ideal-gas picture adopted for the surface pressure. The impurities may originate from the polystyrene substrate, from the surrounding air (our experiments were performed at ambient conditions), or from other unknown sources, despite the care taken in cleaning up all the equipment carefully and despite our use of ultrapure water for the experiments.

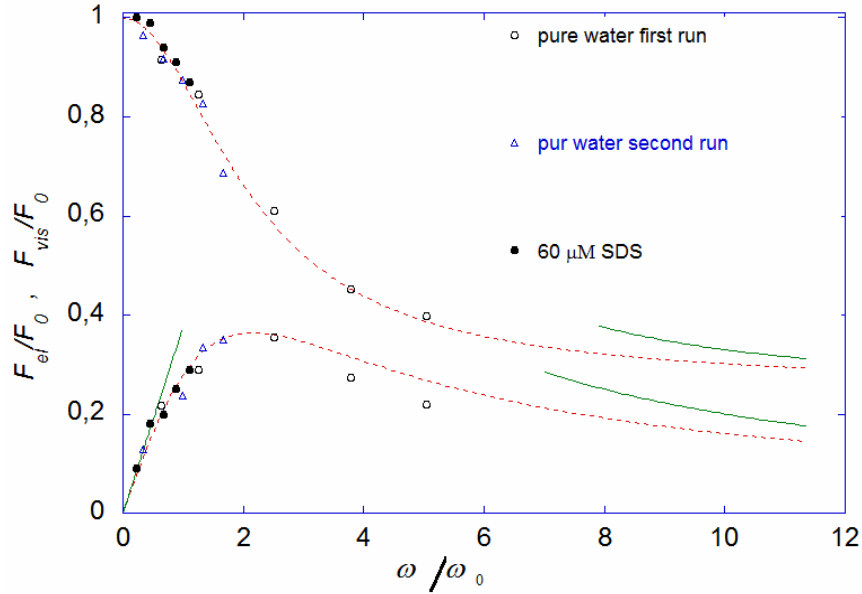

**Fig.5**: Visco-elastic data on "pure" water (first and the second run) and on 60 µM SDS, as a function of the reduced frequency $\omega/\omega_0$. The dashed lines are the fitting curve using numerical calculation, and the continuous ones are given by (10) and (11).

In order to confirm the role of impurities, we have done a control experiment on a 60 μM SDS solution; the viscoelastic response is shown in Fig. 4b, and fitted with a surface pressure $\Pi_0 = 1.0 \pm 0.1$ mN/m. Our surface tension measurements using a Wilhelmy plate, give a tension reduction $\Delta\sigma = 1.2 \pm 0.1$ mN/m. We conclude that this control experiment provides a quantitative confirmation of the above analysis.

In Fig. 5 we plot the measured forces as a function of the reduced frequency $\omega/\omega_0$. The data from the two independent runs and from the control experiment, collapse onto a single curve. The dashed lines are from the numerical calculations and the continuous lines are the analytical results given in Eqs. (10) and (11). The viscous drag force shows a smooth crossover from the no-slip value at zero frequency to the full-slip value at large $\omega$, expected for a free surface. The elastic component increases linearly, passes through a maximum $\omega \approx 2\omega_0$, and then is proportional to $1/\omega$. The analytical calculations describe the asymptotic behavior rather well.

In summary, our experiments demonstrate that we are able to detect impurities at an air-water interface through its visco-elastic response to a vibrating nearby AFM tip. When varying the frequency we observe a cross-over from no-slip to slip boundary conditions. Besides the reduction of the viscous force, we also observe an elastic response, which vanishes in the limits of zero and high frequencies and which is comparable to the viscous drag in the intermediate range. The frequency dependence of both viscous and elastic forces is quantitatively described by our analytical and numerical calculations, with the impurity concentration $c_0$ as the only fit parameter. Our analysis is confirmed by a control experiment where the viscoelastic response results from an added surfactant at known concentration. These results lead us to the the conclusion that very low concentrations of surface impurities (corresponding to an equivalent concentration $10^{-3}$ of the critical micellar concentration of a typical surfactant molecule) drastically modify boundary conditions for flows near interfaces.